\documentclass[12pt,a4paper]{article}
\usepackage{graphicx}

\begin{document}
\textwidth=135mm
 \textheight=200mm
\begin{center}
{\bfseries COMPASS results on
Collins and Sivers asymmetries\\
for charged hadrons
%\footnote{{\small Talk at the Round Table Discussion "Searching for the mixed phase
%of strongly interacting matter at the JINR Nuclotron", JINR, Dubna, July 7 - 9, 2005.}}
}
\vskip 5mm
Anna Martin$^{\ddag}$\\ on behalf of the COMPASS Collaboration
\vskip 5mm
{\small {\it $^\ddag$ Trieste University and INFN, Trieste, Italy}} 
\end{center}
\vskip 5mm
\centerline{\bf Abstract}
The study of transverse spin and transverse momentum effects is an
important part of the scientific program of COMPASS, a fixed target experiment
at the CERN SPS taking data since 2002. 
The studies are carried on by measuring the hadrons produced
in deep inelastic scattering (DIS) of 160 GeV/c muons
off different targets. 
Among the possible asymmetries in the hadron azimuthal distributions, 
particularly interesting are
the Collins and Sivers asymmetries which  the COMPASS Collaboration has 
measured using transversely polarised deuteron and proton targets.
Here new results for charged 
pions and kaons obtained from the 2010 run with a transversely
polarised proton target are presented for the first time.
\vskip 10mm

%\section{\label{sec:intro}Introduction}
In the last ten years, remarkable progress has 
been done in the understanding of
the nucleon structure, and in particular the transverse spin and the transverse
momentum structure~\cite{AidalaBarone}.
Thanks to this worldwide theoretical and experimental effort,
today it is well known that in order to describe the nucleon structure
in the collinear approximation, three parton distribution functions (PDFs)
are needed for the different flavours $q$:
to the two quite well known unpolarised distributions $f_1^q(x)$
and the helicity distributions $g_1^q(x)$ one has to add the transversity
PDFs $h_1^q(x)$, which describe the transverse polarisation of quarks inside
transversely polarised nucleons.
When considering parton intrinsic transverse momentum a total of eight
transverse momentum dependent (TMD) PDFs are needed, the most famous
of them being the Sivers function which describes the correlation
between the parton transverse momentum and the transverse spin
of the parent nucleon.
All these new objects are either poorly known 
(transversity and Sivers functions)
or still unknown and at present the best channel to investigate them
is semi-inclusive DIS (SIDIS) thanks to the
coupling with chiral-odd fragmentation functions (as in the case of
transversity, coped with the Collins function) and/or final state interactions.
Thus in the SIDIS cross-section several modulations in different
combinations of the azimuthal angles of the final state hadron ($\phi_h$) and
of the nucleon spin ($\phi_S$) in the Gamma-Nucleon reference system appear,
and the measurable amplitudes of some of these modulations
are just the convolutions over transverse momenta of transversity or of 
the other TMD PDFs with the fragmentation functions.
Also, when these amplitudes are measured with different targets and
for different identified final state hadrons, the contributions
of quarks of different flavors can be singled out.
 
The Collins asymmetry is the amplitude of the sin$\Phi_C$ modulation
($\Phi_C=\phi_h+\phi_S+\pi$ is the ``Collins angle'') divided by the
transverse nucleon polarisation and the kinematical spin transfer coefficient.
It gives the convolution of transversity and the Collins fragmentation
function on which independent information is coming from measurements at
Belle and BaBar.
Extractions of transversity from the SIDIS and $e^+e^-$ data are being
performed since 2005, when the first results became available.

The Sivers asymmetry is the amplitude of the sin$\Phi_S$ modulation
($\Phi_S=\phi_h-\phi_S$) divided by the
transverse nucleon polarisation and gives
the convolution of the Sivers function and the usual fragmentation
function.
As for transversity, the first extractions were performed in
2005, when the first SIDIS results were produced.

The COMPASS Collaboration has measured all the transverse spin azimuthal 
asymmetries for protons and deuterons~\cite{bakur},
as well as the longitudinal spin~\cite{Alekseev:2010dm} 
and the unpolarised azimuthal asymmetries 
on deuteron~\cite{giulio}.
COMPASS has also measured the transverse spin 
azimuthal asymmetry in hadron pair
production~\cite{christopher}.
In this contribution the most recent COMPASS results on the Collins and the 
Sivers asymmetries are presented.

In COMPASS SIDIS off transversely polarised targets has been measured 
using a 160 GeV $\mu ^+$ beam and a transversely polarised deuteron ($^6$LiD)
target in the years 2002, 2003 and 2004.
A transversely polarised proton (NH$_3$) target was used in 2007 and 2010.
The experimental apparatus was upgraded in 2005 when
the new polarised target magnet with a larger
angular acceptance was installed, and the central region of the RICH detector
was improved.
The data analysis is very similar for all the years and 
the relevant cuts applied to select the 
``standard sample''  are also the same.
In order to ensure the DIS regime, only events with photon virtuality 
$Q^2 > 1$ GeV$^2$, fractional energy of the virtual photon
$ 0.1 < y < 0.9$, and mass of the hadronic final state system 
$W > 5$ GeV/c$^2$ are considered.
The charged hadrons are required to have at least 0.1 GeV/c transverse 
momentum $p_T^h$
with respect to the virtual photon direction and a fraction 
of the available energy $z > 0.2$. 
No momentum cut is applied to measure the asymmetries for charged hadrons.
Charged pions and kaons are identified with
the RICH detector in the momentum range from threshold
(2.7 GeV/c and 9.7 GeV/c respectively) up to 50 GeV/c.
\begin{figure}[tb]
\begin{center}
\includegraphics[trim=0.5cm 0.40cm 0cm 0.40cm, clip=true, width=0.5\textwidth]{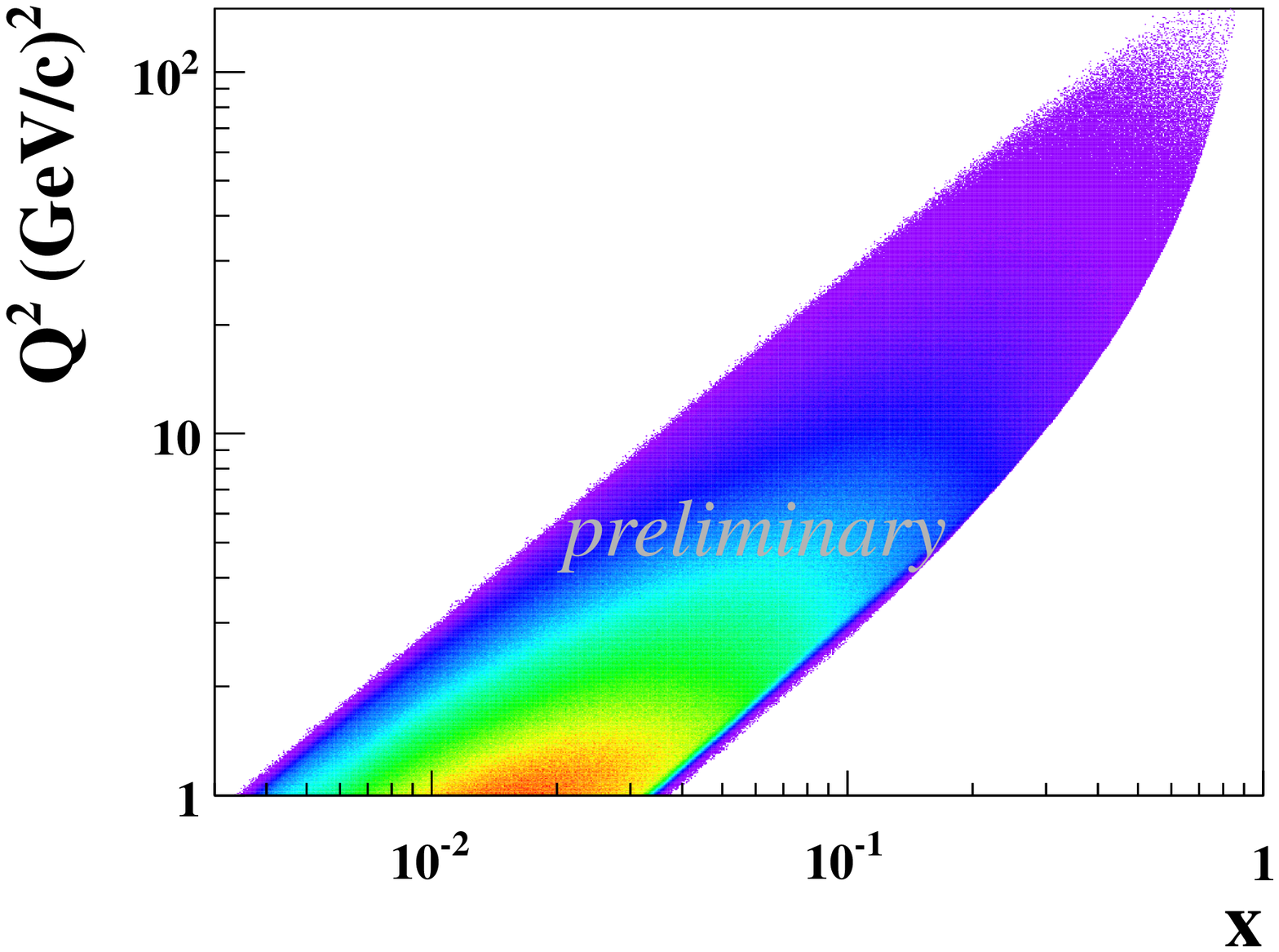}
% l b r t
\includegraphics[trim=0.5cm 0.40cm 0cm 0.40cm, clip=true, width=0.5\textwidth]{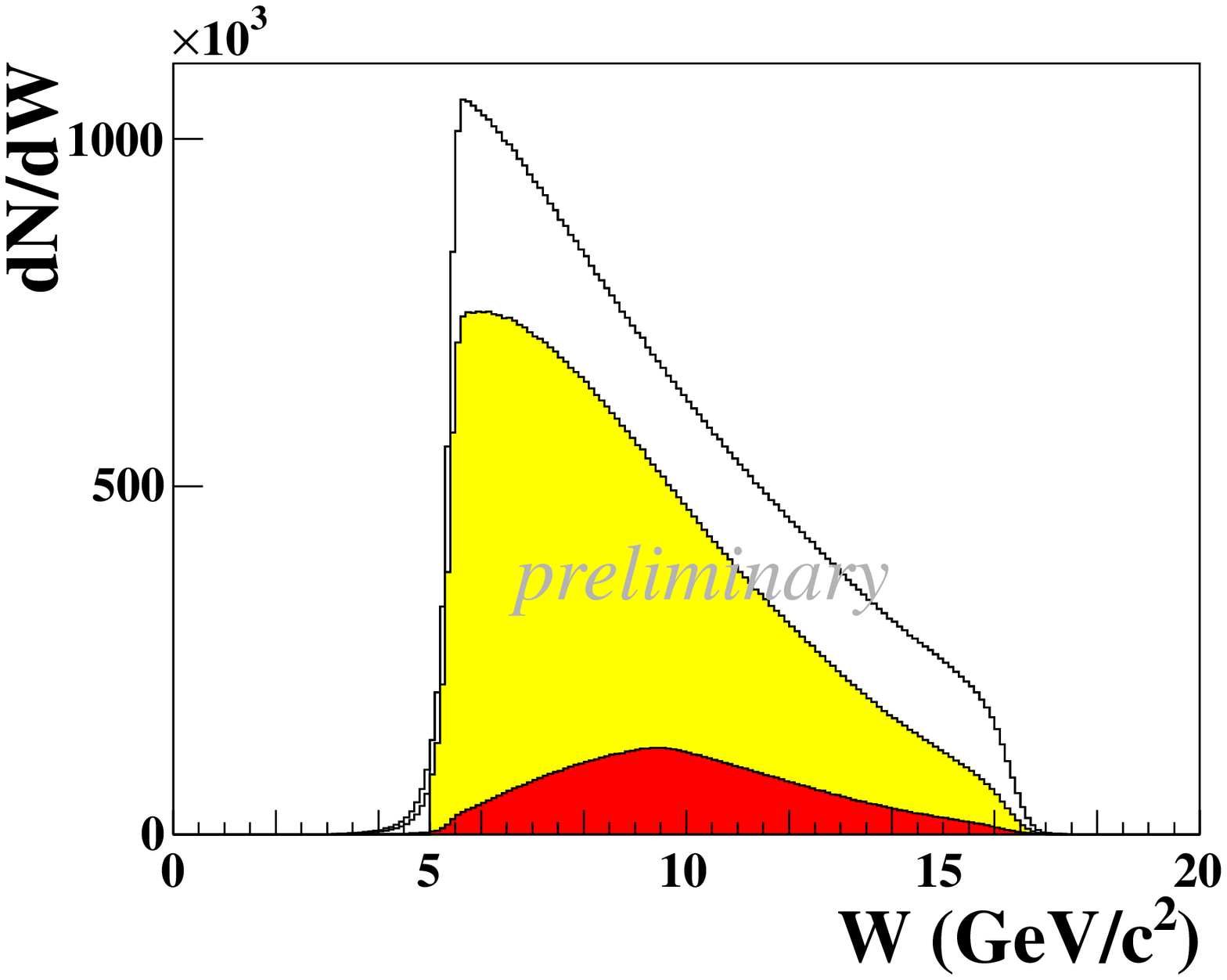}
\vspace*{-.8cm}
 \caption{Left: $x-Q^2$ correlation for charged pions.
Right: $W$ distribution for charged hadrons (upper histogram), pions (middle) 
and kaons (lower).
\label{fig:kin}}
\end{center}
  \end{figure}
The $x-Q^2$ correlation for charged pions from 2010 data
is shown in fig.~\ref{fig:kin} 
(left).
As can be seen, the $x$ range goes from
 $x\simeq 3 \cdot 10^{-3}$ to $x\simeq 0.7$ with relatively large
$Q^2$ values in the valence region.
Fig.~\ref{fig:kin} 
(right) gives the
$W$ distribution for charged hadrons, pions
and kaons.
Here the different shapes are due to the quoted momentum cuts.
The transverse-spin asymmetries are finally 
measured separately for positive and negative hadrons, pions and kaons 
as functions of $x, \, z$ or $p_T^h$ using an unbinned maximum
likelihood method in which all amplitudes are fitted at the same
time.
The correlation between the asymmetries measured as functions of the three 
kinematical
variables, as well as the statistical correlations between the different 
azimuthal asymmetries
measured from the same data, have also been evaluated
and, in the case of the Collins 
and Sivers asymmetries, they have been found to be always smaller than 
0.2~\cite{Adolph:2012sn}.

The Collins and Sivers asymmetries for positive and negative 
hadrons on deuteron~\cite{Alexakhin:2005iw,Ageev:2006da} 
turned out to be compatible with zero
within the few percent uncertainties, at variance
with the non-zero results obtained by the HERMES experiment on 
proton~\cite{Airapetian:2010ds,Airapetian:2009ae}.
Similar results were obtained for charged pions and for 
kaons~\cite{Alekseev:2008aa}.
These data could be understood in terms of a cancellation between 
the $u$ and $d$ quark
contributions with the deuteron target.
Still, the relevance of the different $Q^2$ values in the overlap region
$x>0.03$  of the
 two experiments was an open point, and a COMPASS measurement with
a transversely polarised proton target was urgently needed.

\begin{figure}[tb]
\begin{center}
\includegraphics[trim=0cm 0.20cm 0cm 0.50cm, clip=true, width=0.9\textwidth]{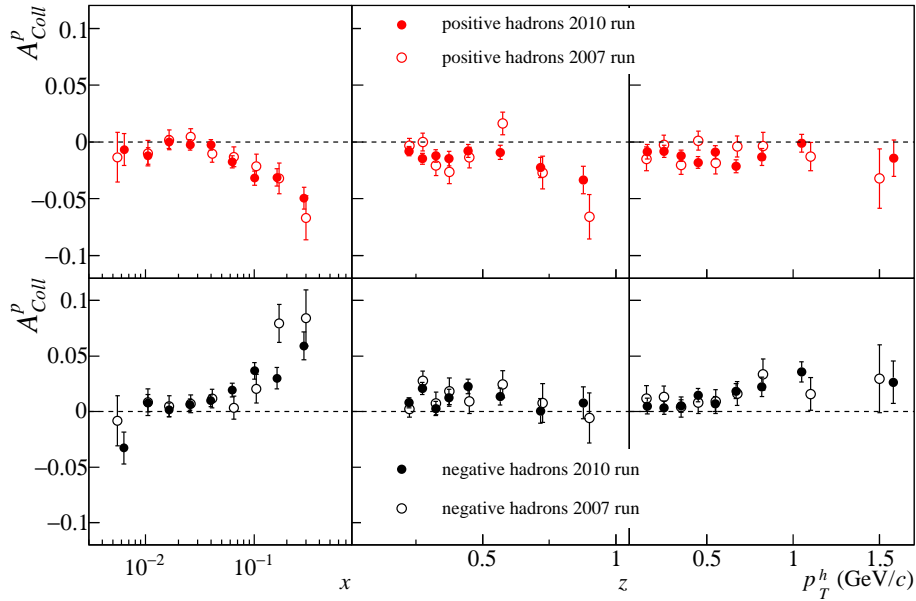}
\vspace*{-.5cm}
 \caption{Collins asymmetries for positive (top) and negative (bottom) hadrons
as functions of $x$, $z$ and $p_T^h$.
The closed (open) points are the results from the 2010 (2007) data. 
In all the figures the error bars show the statistical uncertainties.
The relevant systematic uncertainties are point to point only 
and they are about 0.5 the 
statistical uncertainties.
\label{fig:collinsh}}
\end{center}
  \end{figure}
\begin{figure}[h!]
\begin{center}
\includegraphics[trim=0cm 0.20cm 0cm 0.50cm, clip=true, width=0.9\textwidth]{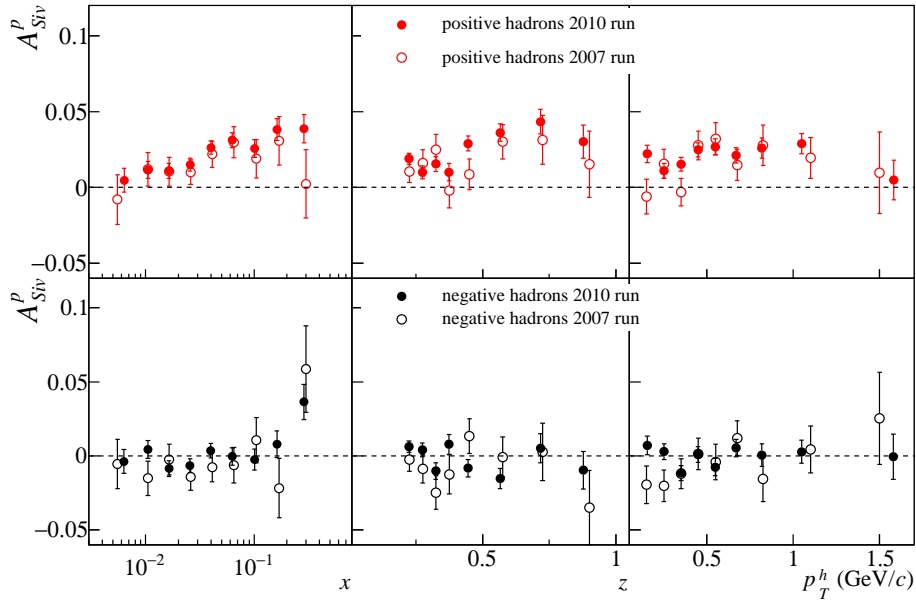}
\vspace*{-.5cm}
 \caption{Same as fig.~\ref{fig:collinsh} for the Sivers asymmetries.
For the 2007 Sivers asymmetry for positive hadrons
only, an overall systematic uncertainty of 0.01 has to be added.
\label{fig:siversh}}
\end{center}
  \end{figure}
The first results on proton from COMPASS came from the analysis
of the 2007 data~\cite{Alekseev:2010rw} and gave strong evidence for non vanishing
effects at COMPASS energies too.
Further data with the transversely polarised 
proton target were taken in a longer run in 2010.
The asymmetries measured from these last 
data~\cite{Adolph:2012sn,Adolph:2012sp}
have statistical and point-to-point systematic 
uncertainties about twice smaller than those of the 2007 data
and their systematic scale uncertainties are negligible
for all the asymmetries.
The results for the Collins asymmetries for positive and negative 
hadrons are compared with the 2007 results in fig.~\ref{fig:collinsh}.
The same comparison for the Sivers asymmetry is shown in 
fig.~\ref{fig:siversh}.
As can be seen the two independent sets of data are in very good agreement.
The Collins asymmetries 
are compatible with zero in the previously unmeasured $x<0.03$ region while 
at larger $x$ they are clearly different from zero, 
with opposite sign for positive and 
negative hadrons and in nice agreement, both in sign and in
magnitude, with the HERMES results~\cite{Airapetian:2010ds}.
The Sivers asymmetry for negative hadrons is compatible 
with zero with some indication for small negative values but 
in the last point, where both measurements give a positive value.
In the case of positive hadrons, the Sivers asymmetry is positive down 
to very small $x$  values and smaller than the same asymmetry for pions
measured by HERMES~\cite{Airapetian:2009ae} in the $x>0.03$ region,
a fact which can be understood in terms of the recent calculations
on TMDs evolution.

\begin{figure}[tb]
\begin{center}
\includegraphics[trim=0cm 0.20cm 0cm 0.80cm, clip=true, width=0.9\textwidth]{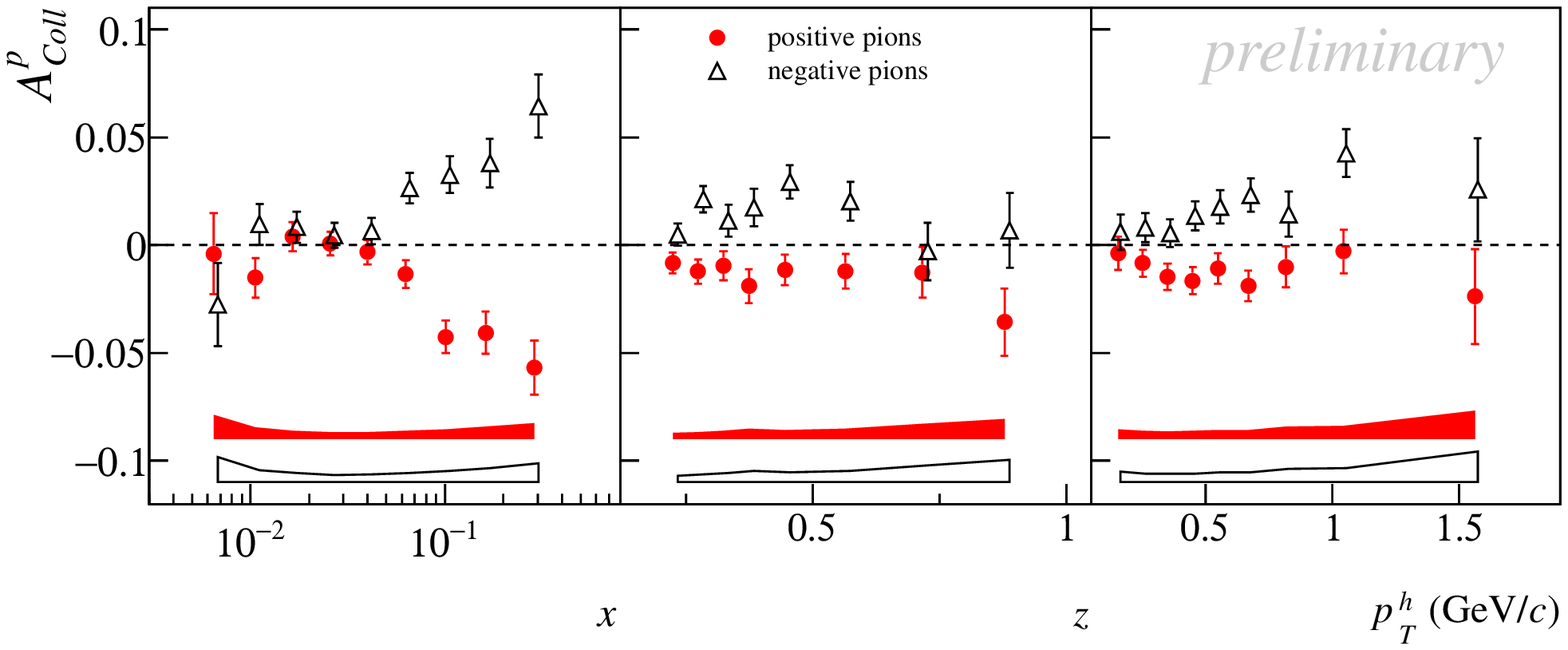}
\includegraphics[trim=0cm 0.20cm 0cm 0.80cm, clip=true, width=0.9\textwidth]{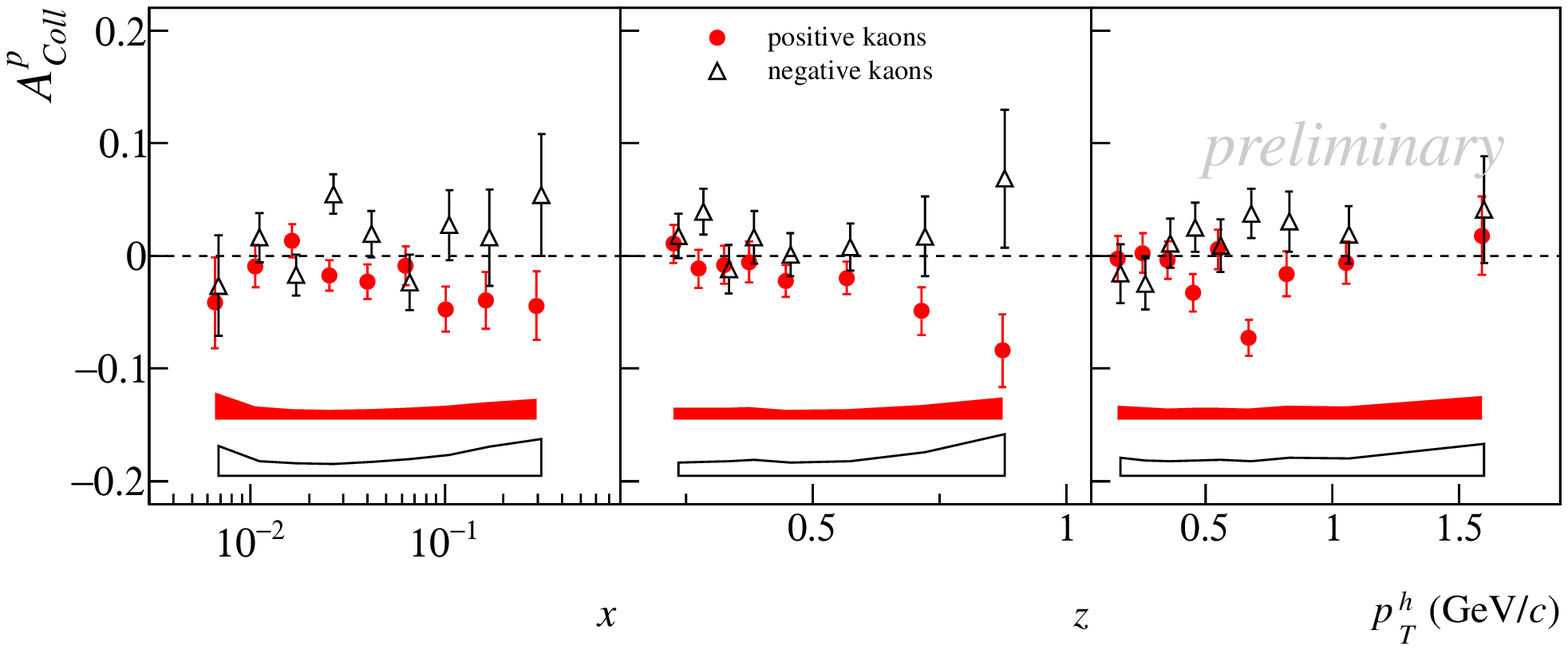}
% l b r t
\vspace*{-.5cm}
 \caption{Collins 
asymmetries for positive (open points) and negative (closed)
pions (upper panel)
and kaons (lower panel) from the 2010
COMPASS proton data.
\label{fig:collinspk}}
\end{center}
  \end{figure}
\begin{figure}[tb]
\begin{center}
\includegraphics[trim=0cm 0.20cm 0cm 0.80cm, clip=true, width=0.9\textwidth]{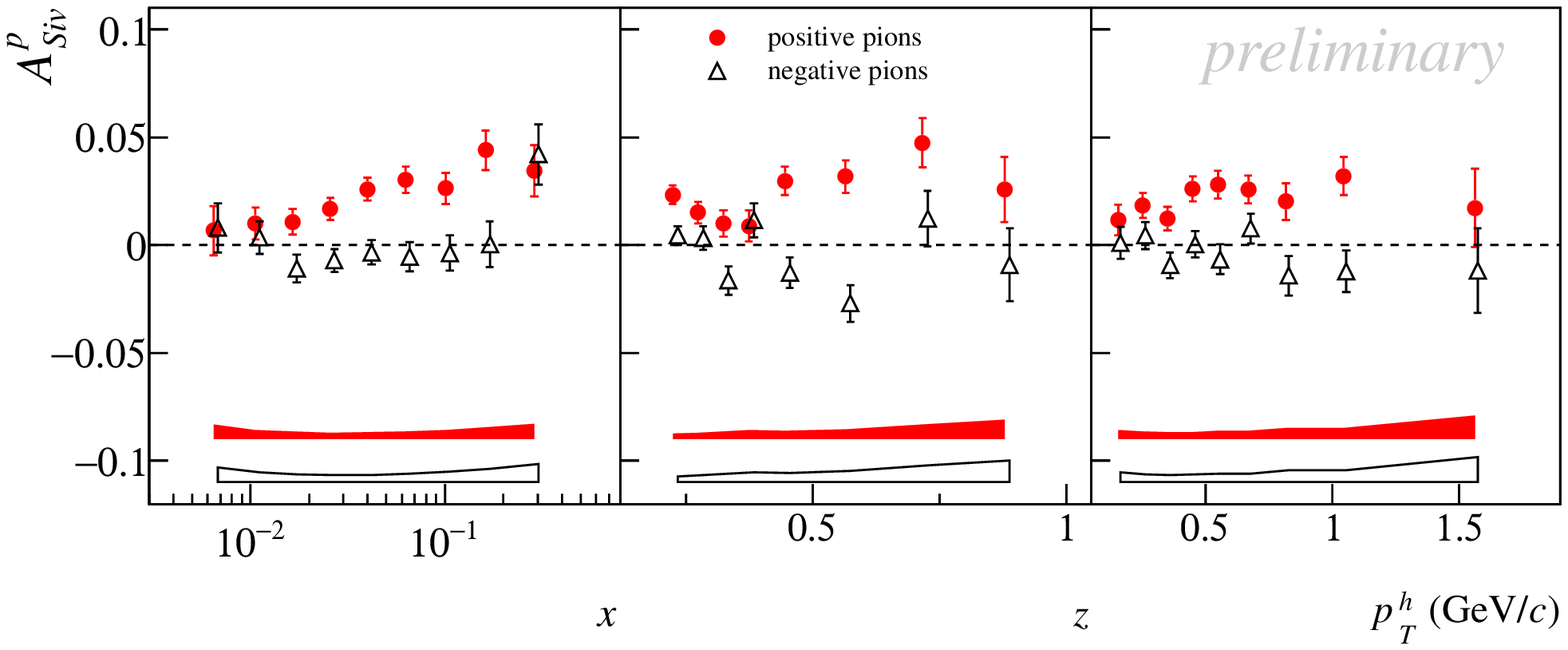}
\includegraphics[trim=0cm 0.20cm 0cm 0.80cm, clip=true, width=0.9\textwidth]{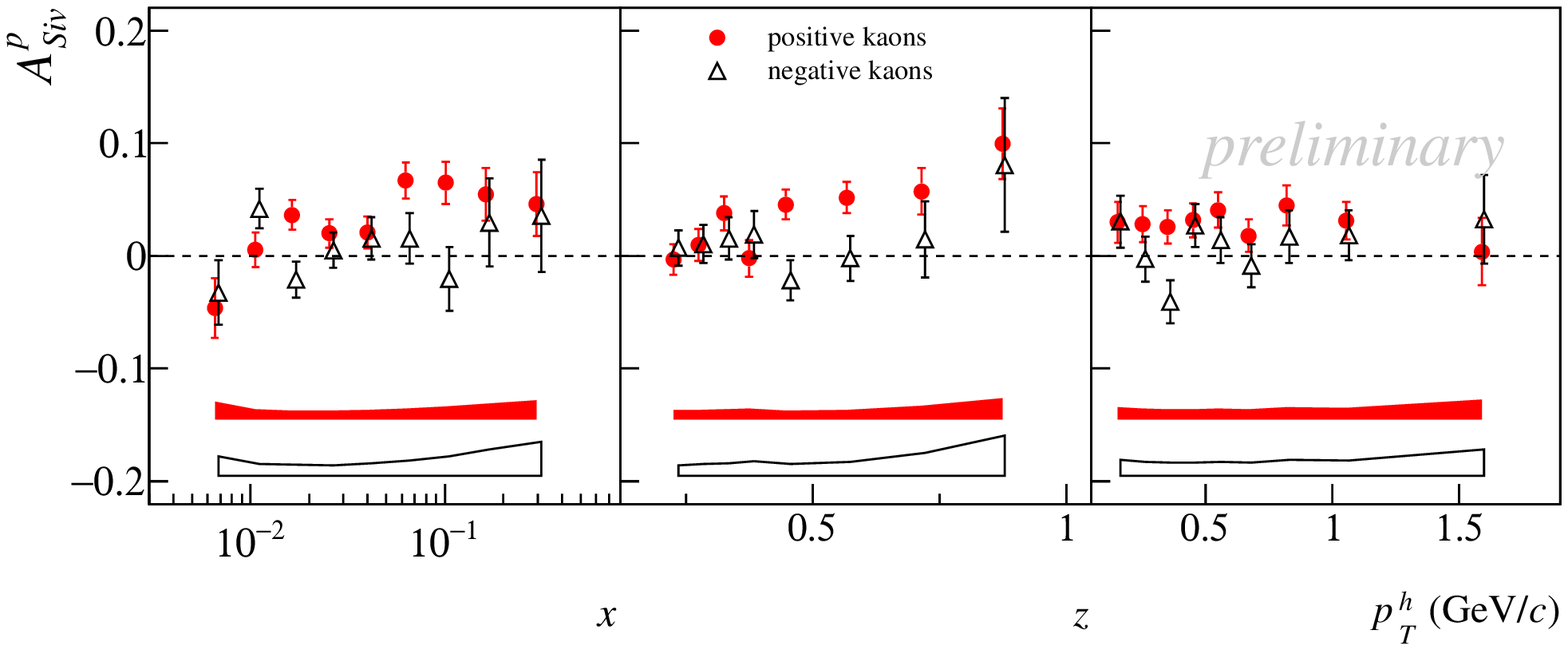}
\vspace*{-.5cm}
 \caption{Sivers
asymmetries for positive (open points) and negative (closed)
pions (upper panel)
and kaons (lower panel) from the 2010
COMPASS proton data.
\label{fig:siverspk}}
\end{center}
  \end{figure}
The Collins and the Sivers asymmetries have also been measured for charged 
pions and kaons.
Preliminary results from the 2007 run
were produced two years ago~\cite{Pesaro:2011zz}.
By now the analysis of the 2010 data is also finished.
As in the case of charged hadrons
the preliminary results from these data
are in very good agreement with the results
from the 2007 data and have smaller statistical and systematic uncertainties. 
The Collins (Sivers) asymmetry from the 2010 data
for charged pions and kaons are 
shown in fig.~\ref{fig:collinspk} (fig.~\ref{fig:siverspk}).

As expected, the values of the pion asymmetries are very close to those of 
the charged hadron asymmetries, and all the previous considerations can be repeated.
In particular, there is no indication for lower values of the Collins asymmetry at the higher COMPASS
$Q^2$ values as compared to the HERMES measurement, while the
Sivers asymmetry for positive pions measured in COMPASS is 
definitively smaller.

For the charged kaons, the statistical uncertainties are larger, still 
the data give some interesting indication.
For the Collins asymmetry there is some indication for different from
zero values.
In the case of the Sivers asymmetry for positive kaons, the values are clearly positive,
again smaller than those measured by HERMES, but higher than those measured 
in COMPASS for positive pions.
A larger  Sivers asymmetry for positive kaons was already observed by HERMES
and it was a puzzling point for the global fits. 
It will be interesting to see which impact
the new COMPASS result will give to a global analysis.

Summarising, precise measurements of the Sivers and Collins asymmetries
on proton have been produced by COMPASS both for charged and identified
final state hadrons.
While no signal was observed in the previous measurements with a deuteron
target, clear non-zero signals are seen for the Collins
and the Sivers asymmetries obtained from the 2007 and 2010 proton data.
These data, when
used together with the results at lower beam energy produced by HERMES and 
the past and future Jefferson Lab experiment, will give
key input for testing the $Q^2$ evolution.
More results for the Collins and Sivers asymmetries 
will come soon from COMPASS.
In particular, thanks to the relatively high statistics of the proton data, 
a multidimensional
analysis to better disentangle the different effects is feasible,
as well as the extraction of the asymmetries in
extended $z$ and $y$ ranges~\cite{Adolph:2012sn,Adolph:2012sp}.
On a longer term, 
the possibility to go back to the measurements with transversely 
polarised targets is under investigation.
More precise measurements with the deuteron target are
needed to better evaluate the
contribution of the $d$ quark, while measurements at lower
energies (e.g. 100 GeV) would allow to better investigate
the energy dependence of the measured effects.

\end{document}